\documentclass[a4paper]{article}
\usepackage[margin=25mm]{geometry}
\usepackage{url}
\usepackage{import}
\usepackage{amsfonts,amssymb,amsmath,amsthm,dsfont}
\usepackage{xcolor}
\usepackage{comment}
\usepackage[linesnumbered,algoruled]{algorithm2e}
\usepackage{bm}
\usepackage[normalem]{ulem}
\usepackage{graphicx}
\graphicspath{{figures/}}
\usepackage{adjustbox}
\usepackage{mathtools}

\providecommand{\keywords}[1]
{
  \small
  \textbf{\textit{Keywords---}} #1
}
\begin{document}

\title{Phase recovery with Bregman divergences for audio source separation\thanks{This work is supported by the European Research Council (ERC FACTORY-CoG-6681839).}}
\date{}
\author{Paul~Magron\thanks{IRIT, Université de Toulouse, CNRS, Toulouse, France (e-mail: firstname.lastname@irit.fr).} ,
        Pierre-Hugo~Vial\footnotemark[2] ,
        Thomas~Oberlin\thanks{ISAE-SUPAERO, Université de Toulouse, France (e-mail: firstname.lastname@isae-supaero.fr).} ,
        Cédric~Févotte\footnotemark[2]
}

\maketitle
\begin{abstract}
Time-frequency audio source separation is usually achieved by estimating the short-time Fourier transform (STFT) magnitude of each source, and then applying a phase recovery algorithm to retrieve time-domain signals. In particular, the multiple input spectrogram inversion (MISI) algorithm has shown good performance in several recent works. This algorithm minimizes a quadratic reconstruction error between magnitude spectrograms. However, this loss does not properly account for some perceptual properties of audio, and alternative discrepancy measures such as beta-divergences have been preferred in many settings. In this paper, we propose to reformulate phase recovery in audio source separation as a minimization problem involving Bregman divergences. To optimize the resulting objective, we derive a projected gradient descent algorithm. Experiments conducted on a speech enhancement task show that this approach outperforms MISI for several alternative losses, which highlights their relevance for audio source separation applications.
\end{abstract}
\keywords{Phase recovery, Bregman divergences, projected gradient descent, audio source separation, speech enhancement.}
\section{Introduction}
\label{sec:intro}

Audio source separation~\cite{Comon2010} consists in extracting the underlying \textit{sources} that add up to form an observable audio \textit{mixture}. This task finds applications in many areas such as speech enhancement and recognition~\cite{Barker2018} or musical signal processing~\cite{Cano2019}. State-of-the-art approaches for source separation consist in using a deep neural network (DNN) or nonnegative matrix factorization (NMF) to estimate a nonnegative mask that is applied to a time-frequency (TF) representation of the audio mixture, such as the short-time Fourier transform (STFT)~\cite{Wang2018c}. Recent works such as~\cite{Luo2019,Luo2020} operate in the time domain directly, but TF approaches remain interesting since they allow to better exploit the structure of sound~\cite{Ditter2020}.

Applying a nonnegative mask to the mixture's STFT results in assigning its phase to each isolated source. Even though this practice is common and yields satisfactory results, it is well established~\cite{Magron2018b} that when sources overlap in the TF domain, using the mixture's phase induces residual interference and artifacts in the estimates. With the advent of deep learning, magnitudes can nowadays be estimated with a high accuracy, which outlines the need for more advanced phase recovery algorithms~\cite{Gerkmann2015}. Consequently, a significant research effort has been put on phase recovery in DNN-based source separation, whether phase recovery algorithms are applied as a post-processing~\cite{Magron2018b} or integrated within end-to-end systems for time-domain separation~\cite{Wang2018a,Wichern2018,Wisdom2019}.

Among the variety of phase recovery techniques, the multiple input spectrogram inversion (MISI) algorithm~\cite{Gunawan2010} is particularly popular. This iterative procedure consists in retrieving time-domain sources from their STFT magnitudes while respecting a mixing constraint: the estimates must add up to the mixture. This algorithm exhibits a good performance in source separation when combined with DNNs~\cite{Wang2018a,Wichern2018}. However, MISI suffers from one limitation. Indeed, it is derived as a solution to an optimization problem that involves the quadratic loss, which is not the best-suited metric for evaluating discrepancies in the TF domain. For instance, it does not properly account for the large dynamic range of audio signals~\cite{Gray1980}.

In this work, we consider phase recovery in audio source separation as an optimization problem involving  alternative divergences which are more appropriate for audio processing. We consider general Bregman divergences, a family of loss functions which encompasses the beta-divergence~\cite{Hennequin2011} and some of its well-known special cases, the Kullback-Leibler (KL) and Itakura-Saito (IS) divergences. These divergences are acknowledged for their superior performance in audio spectral decomposition applications such as NMF-based source separation~\cite{Smaragdis2014}. In a previous work~\cite{Vial2021}, we addressed phase recovery with the Bregman divergences in a single-source setting. Here, we propose to extend this approach to a single-channel and multiple-sources framework, where the mixture's information can be exploited. To optimize the resulting objective, we derive a projected gradient algorithm~\cite{Combettes2011}. We experimentally assess the potential of our approach for a speech enhancement task. Our results show that this method outperforms MISI for several Bregman divergences.

The rest of this paper is structured as follows. Section~\ref{sec:related_work} presents the related work. In Section~\ref{sec:method} we derive the proposed algorithm. Section~\ref{sec:exp} presents the experimental results. Finally, Section~\ref{sec:conc} draws some concluding remarks.

\vspace{0.5em}
\noindent \textbf{Mathematical notations}:
\begin{itemize}
    \item $\mathbf{A}$ (capital, bold font): matrix.
    \item $\mathbf{s}$ (lower case, bold font): vector.
    \item $\text{diag}(\mathbf{u}) \in \mathbb{C}^{K \times K}$: diagonal matrix whose entries are the elements of $\mathbf{u} \in \mathbb{C}^{K}$.
    \item $z$ (regular): scalar.
    \item $|.|$, $\angle(.)$: magnitude and complex angle, respectively.
    \item $\mathbf{s}^\mathsf{T}$, $\mathbf{s}^\mathsf{H}$: transpose and Hermitian transpose, respectively.
    \item $\Re(.)$, $\Im(.)$: real and imaginary part functions, respectively.
    \item $||.||_2$: Euclidean norm.
    \item $\odot$, $(.)^d$, fraction bar: element-wise matrix or vector multiplication, power, and division, respectively.
\end{itemize}

\section{Related work}
\label{sec:related_work}

In this section, we present the necessary background upon which our work builds. We describe the baseline phase recovery problem (Section~\ref{sec:related_work_GL}), its extension to multiple sources (Section~\ref{sec:related_work_misi}), and its formulation using the Bregman divergences (Section~\ref{sec:related_work_breg}).

\subsection{Phase recovery}
\label{sec:related_work_GL}

Phase recovery is commonly formulated as the following problem:
\begin{equation}
    \label{eq:pr}
    \underset{\mathbf s \in \mathbb R^L}{\text{min}} \|\mathbf r - |\mathbf A \mathbf s|^d\|_2^2,
\end{equation}
where $\mathbf r \in \mathbb{R}_+^K$ are nonnegative measurements, usually an STFT magnitude ($d=1$) or power ($d=2$) spectrogram, and $\mathbf A \in \mathbb{C}^{K \times L}$ is the matrix that encodes the STFT. In the seminal work~\cite{Griffin1984}, the authors address problem~\eqref{eq:pr} with $d=1$. Starting from an initial guess $\mathbf s^{(0)}$, they propose the following update rule:
\begin{equation}
    \label{eq:gl}
    \mathbf s^{(t+1)}  = \mathbf A^\dag \left( \mathbf r \odot \frac{\mathbf A \mathbf s^{(t)}}{|\mathbf A \mathbf s^{(t)}|} \right)
\end{equation}
where $\mathbf A^\dag$ is the Moore-Penrose pseudo-inverse of $\mathbf A$ defined as $\mathbf A^\dag = (\mathbf A^\mathsf H \mathbf A)^{.-1} \mathbf A^\mathsf H$, which encodes the inverse STFT.
This iterative scheme, known as the Griffin-Lim (GL) algorithm, is proved to converge to a critical point of the quadratic loss in \eqref{eq:pr}~\cite{Griffin1984}, and can also be obtained by majorization-minimization~\cite{Qiu2016} or using a gradient descent scheme~\cite{Vial2021}. Improvements of this algorithm notably include accelerated~\cite{Perraudin2013} and real-time purposed versions~\cite{Zhu2007}.

\subsection{Multiple input spectrogram inversion (MISI)}
\label{sec:related_work_misi}

The GL algorithm has been extended to handle multiple sources in a source separation framework~\cite{Gunawan2010}. Given an observed mixture $\mathbf x \in \mathbb{R}^L$ of $C$ sources $\mathbf s_c \in \mathbb{R}^L$, whose target nonnegative TF measurements are $\mathbf r_c$, this problem can be formulated as~\cite{Magron2020}:
\begin{equation}
    \label{eq:misi_prob}
    \underset{ \{ \mathbf s_c \in \mathbb R^L \}_{c=1}^C }{\text{min}} \sum_{c=1}^C \|\mathbf r_c - |\mathbf A \mathbf s_c|^d\|_2^2 \text{ s.t. } \sum_{c=1}^C \mathbf s_c = \mathbf x.
\end{equation}
The multiple input spectrogram inversion (MISI) algorithm addresses~\eqref{eq:misi_prob} when $d=1$ and consists of the following updates:
\begin{align}
    \forall c \text{, } \mathbf y_c^{(t)}  &= \mathbf A^\dag \left( \mathbf r \odot \frac{\mathbf A \mathbf s_c^{(t)}}{|\mathbf A \mathbf s_c^{(t)}|} \right) \label{eq:misi_updates_mag} \\
    \forall c \text{, } \mathbf s_c^{(t+1)} &=  \mathbf y_c^{(t)} + \frac{1}{C} \left( \mathbf x - \sum_{i=1}^C \mathbf y_i^{(t)} \right) \label{eq:misi_updates_proj}
\end{align}
In a nutshell, this algorithm consists in performing the GL update~\eqref{eq:gl} for each source individually, and then distributing the resulting mixing error onto those estimates to yield a set of signals $\{\mathbf s_c\}$  that add up to the mixture. The MISI algorithm has been introduced heuristically in~\cite{Gunawan2010}. In~\cite{Magron2020}, it was derived using a majorization-minimization strategy, which proved its convergence.

\subsection{Phase recovery with the Bregman divergence}
\label{sec:related_work_breg}

In~\cite{Vial2021}, we proposed to replace the quadratic loss in problem~\eqref{eq:pr} with Bregman divergences, which encompass the beta-divergence~\cite{Hennequin2011} and its special cases, the KL and IS divergences. A Bregman divergence $\mathcal D_\psi$ is defined from a strictly-convex, continuously-differentiable generating function $\psi$ (with derivative $\psi'$) as follows:
\begin{equation}
\mathcal D_\psi (\mathbf p\, \bm | \, \mathbf q)= \sum_{k=1}^K \left[ \psi(p_k) - \psi(q_k) - \psi'(q_k)(p_k - q_k) \right].
\label{eq:breg}
\end{equation}
Typical Bregman divergences with their generating function and derivative can be found, e.g., in~\cite{Vial2021} (see Table 1). Since the Bregman divergences are not symmetric in general, we considered the following  two problems, respectively termed ``left" and ``right":
\begin{equation}
    \underset{\mathbf s \in \mathbb R^L}{\text{min}}\, \mathcal D_\psi(\mathbf r \,\bm |\, |\mathbf A \mathbf s|^d) \text{ and } \underset{\mathbf s \in \mathbb R^L}{\text{min}}\, \mathcal D_\psi(|\mathbf A \mathbf s|^d \,\bm |\, \mathbf r).
    \label{eq:bregpr_right}
\end{equation}
We derived two algorithms for solving these, based on gradient descent and alternating direction method of multipliers (ADMM)~\cite{Vial2021}.

\section{Proposed method}
\label{sec:method}

\subsection{Problem setting}

We propose to extend our previous approach described in Section~\ref{sec:related_work_breg} to a single-channel source  separation framework. Indeed, as described in Section~\ref{sec:related_work_misi}, it is necessary to include the mixture information in the optimization problem so that the estimates add up to the mixture. We replace the loss in~\eqref{eq:misi_prob} with a Bregman divergence, as in~\eqref{eq:bregpr_right}, which yields the following optimization problem:
%\footnote{At the time of submission of this paper, another article addressing a similar problem has been published online~\cite{Masuyama2020}. However, this approach, based on an ADMM scheme, is limited to the KL (``right" and ``left") and IS (``left") divergences only, since it involves deriving a proximal operator which is not available in closed-form for general Bregman divergences. As such, we did not include it in our experiments.}
%
\begin{equation}
    \label{eq:bregmisi_prob}
    \underset{ \{ \mathbf s_c \in \mathbb R^L \}_{c=1}^C }{\text{min}} \sum_{c=1}^C J_c(\mathbf{s}_c) \ \text{ s.t. } \sum_{c=1}^C \mathbf s_c = \mathbf x,
\end{equation}
where $J_c(\mathbf{s}_c) = \mathcal D_\psi(\mathbf r_c \,\bm |\, |\mathbf A \mathbf s_c|^d)$ for the ``right" problem and $J_c(\mathbf{s}_c) = \mathcal D_\psi(|\mathbf A \mathbf s_c|^d \,\bm |\, \mathbf r_c )$ for its ``left" counterpart.

\subsection{Projected gradient descent}

Similarly to \cite{Vial2021}, we propose a gradient descent algorithm to minimize the objective defined in \eqref{eq:bregmisi_prob}. The set of signals whose sum is equal to the observed mixture $\mathbf{x}$, appearing in the constraint of~\eqref{eq:bregmisi_prob}, is convex. As such, we may use the projected gradient algorithm~\cite{Combettes2011} which boils down to alternating the two following updates:
\begin{align}
    \forall c \text{, } \mathbf y_c^{(t)}   &= \mathbf s_c^{(t)}  - \mu \nabla J_c ( \mathbf s_c^{(t)} ) \label{eq:bregmisi_updates_grad}  \\
    \forall c \text{, } \mathbf s_c^{(t+1)}  &=  \mathbf y_c^{(t)}  + \frac{1}{C} \left( \mathbf x - \sum_{i=1}^C \mathbf y_i^{(t)}  \right) \label{eq:bregmisi_updates_proj}
\end{align}
where $\nabla J_c$ denotes the gradient of $J_c$ with respect to $\mathbf{s}_c$ and $\mu>0$ is the gradient step size. In a nutshell,~\eqref{eq:bregmisi_updates_grad} performs a gradient descent, and, similarly to~\eqref{eq:misi_updates_proj},~\eqref{eq:bregmisi_updates_proj} projects the auxiliary variables $\mathbf{y}_c$ onto the set of estimates whose sum is equal to the mixture.

\subsection{Derivation of the gradient}

We derive hereafter the gradient of $J_c$. Using the chain rule~\cite{Magnus1985}, we have:
\begin{equation}
    \nabla J_c( \mathbf s_c) =  (\nabla |\mathbf{A} \mathbf{s}_c  |^d)^\mathsf{T} \mathbf{g}_c,
    \label{eq:nabla_Jc}
\end{equation}
where $\nabla |\mathbf{A} \mathbf{s}_c   |^d$ denotes the Jacobian of the multivariate function $\mathbf{s}_c  \to |\mathbf{A} \mathbf{s}_c   |^d$ (the Jacobian being the extension of the gradient for multivariate functions, we may use the same notation $\nabla$), and:
\begin{align*}
   \text{for the ``right" problem, } \mathbf{g}_c &= \psi''(|\mathbf A  \mathbf{s}_c|^d) \odot
    (|\mathbf A  \mathbf{s}_c|^d-\mathbf r_c) \\
    \text{for the ``left" problem, } \mathbf{g}_c &= \psi' (|\mathbf A  \mathbf{s}_c|^d) - \psi'(\mathbf r_c)
\end{align*}
where $\psi'$ and $\psi''$ are applied entrywise. Now, let us note $\mathbf{A}_r$ and $\mathbf{A}_i$ the real and imaginary parts of $\mathbf{A}$, respectively. Using differentiation rules for element-wise matrix operations~\cite{Magnus1985} and calculations similar to~\cite{Vial2021}, we have:
\begin{equation}
\begin{aligned}
    \nabla |\mathbf{A} \mathbf{s}_c  |^d &= \nabla  \left(  (\mathbf{A}_r \mathbf{s}_c)^2 + (\mathbf{A}_i \mathbf{s}_c)^2  \right)^{\frac{d}{2}} \\
    &= d \times \text{diag}( |\mathbf{A} \mathbf{s}_c|^{d-2} )  \left( \text{diag}( \mathbf{A}_r \mathbf{s}_c )  \mathbf{A}_r + \text{diag}( \mathbf{A}_i \mathbf{s}_c )  \mathbf{A}_i  \right). 
\end{aligned}
\label{eq:grad_Axd}
\end{equation}
We now inject~\eqref{eq:grad_Axd} in~\eqref{eq:nabla_Jc} and develop, which yields:
\begin{equation}
    \nabla J_c( \mathbf{s}_c) =   \mathbf{A}_r^\mathsf{T} \left( d \times \text{diag}( \mathbf{A}_r \mathbf{s}_c )  \text{diag}( |\mathbf{A} \mathbf{s}_c|^{d-2} ) \mathbf{g}_c  \right) + \mathbf{A}_i^\mathsf{T} \left( d \times \text{diag}( \mathbf{A}_i \mathbf{s}_c )  \text{diag}( |\mathbf{A} \mathbf{s}_c|^{d-2})\mathbf{g}_c \right).
\end{equation}
We remark that $\forall \mathbf{u}, \mathbf{v} \in \mathbb{C}^K$, $\text{diag}(\mathbf{u}) \mathbf{v} = \mathbf{u} \odot \mathbf{v}$, so we further simplify this expression:
\begin{equation}
    \nabla J_c(  \mathbf{s}_c) =   \mathbf{A}_r^\mathsf{T} \left( d \times (\mathbf{A}_r  \mathbf{s}_c) \odot |\mathbf{A}  \mathbf{s}_c|^{d-2} \odot \mathbf{g}_c  \right) +
    \mathbf{A}_i^\mathsf{T} \left( d \times (\mathbf{A}_i  \mathbf{s}_c) \odot |\mathbf{A}  \mathbf{s}_c|^{d-2} \odot \mathbf{g}_c \right).
    \label{eq:grad_Jc_aux}
\end{equation}
Finally, we remark that $\forall \mathbf{u} \in \mathbb{C}^K$, $\Re (\mathbf{A}^\mathsf{H} \mathbf{u}) = \mathbf{A}_r^\mathsf{T} \Re(\mathbf{u}) + \mathbf{A}_i^\mathsf{T} \Im (\mathbf{u})$, thus we can rewrite the gradient~\eqref{eq:grad_Jc_aux} as:\footnote{Note that this gradient is not defined when at least one entry of $\mathbf{A} \mathbf{s}_c$ is null for $d<2$ and/or $\beta \leq 1$. To alleviate this potential issue, we consider a regularized loss using an additive small value $\varepsilon \ll 1$, as detailed in~\cite{Vial2021}.}
\begin{equation}
    \nabla J_c(  \mathbf{s}_c) = d \times \Re \left( \mathbf{A}^\mathsf{H} ( (\mathbf{A} \mathbf{s}_c) \odot |\mathbf{A} \mathbf{s}_c|^{d-2} \odot \mathbf{g}_c ) \right).
    \label{eq:grad_Jc}
\end{equation}

\subsection{Implementation of the gradient update}

It is common practice to use the same window for computing both the STFT and its inverse, up to a normalization constant $b$, which ensures perfect reconstruction for usual windows (e.g., Hann or Hamming) and overlap ratios (e.g., $50$ or $75$ $\%$)~\cite{Smith2011}. In such a setting, $\mathbf{A}^\mathsf{H} \mathbf{A} = b \mathbf{I}$, and thus $\mathbf{A}^\dag = \frac{1}{b} \mathbf{A}^\mathsf{H}$: consequently, $\mathbf{A}^\mathsf{H}$ encodes the inverse STFT up to this normalization constant.

\noindent Let us also point out that when processing audio signals, applying $\mathbf{A}^\mathsf{H}$ returns real-valued signals~\cite{Vial2021}. We can therefore ignore the extra real part in~\eqref{eq:grad_Jc}. The gradient update~\eqref{eq:bregmisi_updates_grad} then rewrites:
\begin{equation}
    \forall c \text{, }\mathbf y_c^{(t)} = \mathbf s_c^{(t)}  - \tilde{\mu} d \times \left( \mathbf{A}^\dag ( (\mathbf{A} \mathbf{s}_c) \odot |\mathbf{A} \mathbf{s}_c|^{d-2} \odot \mathbf{g}_c ) \right),
    \label{eq:bregmisi_updates_grad_simp}
\end{equation}
where $\tilde{\mu}=\mu / b$ is the normalized step size, which we simply term ``step size" in what follows.

\noindent \textit{Remark}: When considering the quadratic loss (for which the ``right" and ``left" problems are equivalent) with $d=1$ and step size $\tilde{\mu}=1$, the gradient update~\eqref{eq:bregmisi_updates_grad_simp} becomes equivalent to the MISI update~\eqref{eq:misi_updates_mag}. This outlines that our method generalizes MISI, as the latter can be seen as a particular case of the projected gradient descent algorithm.

\subsection{Algorithm}

\begin{algorithm}[t]
	\caption{Phase recovery with the Bregman divergence for audio source separation: gradient descent.}
	\label{al:bregmisi_gradient}
			\textbf{Inputs}: Measurements $\mathbf{R}_c \in \mathbb{R}_+^{M \times N}$, mixture $\mathbf{x} \in \mathbb{R}^{L}$, step size $\tilde{\mu}>0$, Bregman divergence function $\psi$. \\

            \textbf{Initialization}: \\
            $\forall c$, $\mathbf{s}_c = \text{iSTFT}(\mathbf{R}_c^{1/d} \odot \frac{\text{STFT}(\mathbf{x})}{|\text{STFT}(\mathbf{x})|})$ \\
            
			\While{stopping criteria not reached}{
			
			$\forall c$, $\mathbf S_c = \text{STFT}(\mathbf{s}_c)$ \\
			 \uIf{``right"}{
                $\mathbf G_c =  \psi'' (|\mathbf{S}_c|^d) \odot  ( |\mathbf{S}_c|^d -\mathbf{R}_c)$
              }
			\uElseIf{``left"}{
                $\mathbf G_c = \psi' (|\mathbf{S}_c|^d) - \psi' (\mathbf{R}_c)$
              }
            $\forall c$, $\mathbf y_c = \mathbf s_c - \tilde{\mu} d \times \text{iSTFT}(\displaystyle \mathbf S_c \odot |\mathbf S_c|^{d-2} \odot \mathbf G_c )$ \\
            $\forall c$, $\mathbf s_c = \mathbf y_c + ( \mathbf x - \sum_{i=1}^C \mathbf y_i ) / C$
			}
			
			\textbf{Output}: $\{ \mathbf{s}_c \}_{c=1}^C$
		
\end{algorithm}

The proposed algorithm consists of alternating the updates~\eqref{eq:bregmisi_updates_grad_simp} and~\eqref{eq:bregmisi_updates_proj}. A natural choice for obtaining initial source estimates consists in assigning the mixture's phase to each source's STFT, which is known as \emph{amplitude masking} and is commonly employed to initialize MISI~\cite{Wang2018a,Wichern2018,Gunawan2010}:
\begin{equation}
    \forall c \text{, }  \mathbf{s}_c^{(0)} = \mathbf A^\dag \left( \mathbf{r}_c^{1/d} \odot \frac{\mathbf A \mathbf x}{|\mathbf A \mathbf x| } \right).
    \label{eq:am_ini}
\end{equation}
The STFT matrix $\mathbf{A}$ and its inverse are large structured matrices that allow for efficient implementations of matrix-vector products. In that setting, it is more customary to handle TF matrices of size $M \times N$, where $M$ is the number of frequency channels and $N$ the number of time frames, rather than vectors of size $K=MN$. As such, we provide in Algorithm~\ref{al:bregmisi_gradient} the pseudo-code for practical implementation of our method.

\section{Experiments}
\label{sec:exp}

In this section, we assess the potential of Algorithm~\ref{al:bregmisi_gradient} for a speech enhancement task, that is, with $C=2$ and where $\mathbf{s}_1$ and $\mathbf{s}_2$ correspond to the clean speech and noise, respectively. Note however that this framework is applicable to alternative separation scenarios, such as musical instruments~\cite{Cano2019} or multiple-speakers~\cite{Wang2018c} separation. The code related to these experiments is available online.\footnote{\url{https://github.com/magronp/bregmisi}}

\subsection{Protocol}

\paragraph*{Data.}
As acoustic material, we build a set of mixtures of clean speech and noise. The clean speech is obtained from the VoiceBank test set~\cite{Valentini2017}, from which we randomly select $100$ utterances. The noise signals are obtained from the DEMAND dataset~\cite{Thiemann2013}, from which we select noises from three real-world environments: a living room, a bus, and a public square. For each clean speech signal, we randomly select a noise excerpt cropped at the same length than that of the speech signal. We then mix the two signals at various input signal-to-noise ratios (iSNRs) ($10$, $0$, and $-10$ dB). All audio excerpts are single-channel and sampled at $16,000$ Hz. The STFT is computed with a $1024$ samples-long ($64$ ms) Hann window, no zero-padding, and 75$\%$ overlap. The dataset is split into two subsets of $50$ mixtures: a \emph{validation} set, on which the step size is tuned (see Section~\ref{sec:exp_stepsize}); and a \emph{test} set, on which the proposed algorithm is compared to MISI.

\paragraph*{Spectrogram estimation.}
In realistic scenarios, the nonnegative measurements $\mathbf{r}_c$ are estimates of the magnitude or power spectrograms of the sources. To obtain such estimates, we use Open-Unmix~\cite{Stoter2019}, an open implementation of a three-layer BLSTM neural network, originally tailored for music source separation applications. This network has been adapted to a speech enhancement task. It was trained on our dataset, except using different speakers and noise environments, as described in~\cite{Valentini2016}. We use the trained model available at~\cite{Uhlich2020}. This network is fed with the noisy mixtures and outputs an estimate for the clean speech and noise spectrograms, which serve as inputs to the phase retrieval methods.

\paragraph*{Compared methods.}
We test the proposed projected gradient descent method described in Algorithm~\ref{al:bregmisi_gradient} in a variety of settings. We consider magnitude and power measurements ($d=1$ or $2$), ``right" and ``left" problems, and various values of $\beta$ for the divergence ($\beta=0$ to $2$ with a step of $0.25$). The step size is tuned on the validation set. As comparison baseline, we consider the MISI algorithm (which corresponds to our algorithm with $\beta=2$, $d=1$ and $\tilde{\mu}=1$). Following traditional practice with MISI~\cite{Wang2018a,Wichern2018}, all algorithms are run with $5$ iterations.

\noindent In order to evaluate the speech enhancement quality, we compute the signal-to-distortion ratio (SDR) between the true clean speech $\mathbf{s}_1^\star$ and its estimate $\mathbf{s}_1$ (higher is better):
\begin{equation}
    \text{SDR}(\mathbf{s}_1^\star, \mathbf{s}_1) = 20 \log_{10}\frac{\|\mathbf s_1^\star\|_2}{\|\mathbf s_1^\star - \mathbf s_1\|_2}.
\end{equation}
For more clarity, we will present the SDR improvement (SDRi) of a method (whether MISI or Algorithm~\ref{al:bregmisi_gradient}) over initialization.

\subsection{Influence of the step size}
\label{sec:exp_stepsize}

\begin{figure}[t]
    \centering
    \includegraphics[width=0.8\linewidth]{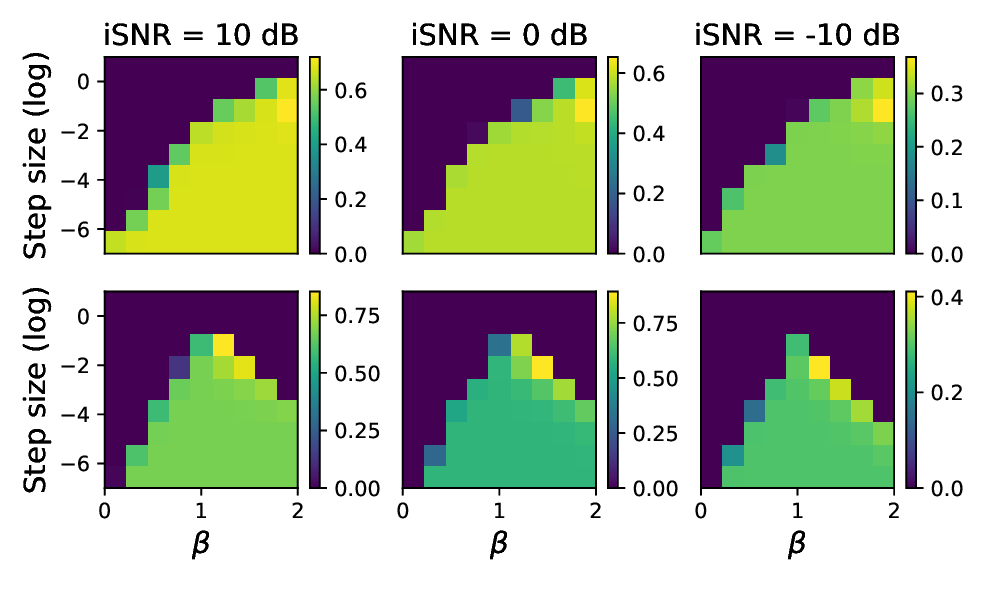}
    \caption{Average SDRi on the validation set obtained with the proposed algorithm at various iSNRs, when $d=1$ (top) and $d=2$ (bottom). For better readability, we set the SDRi at $0$ when convergence issues occur as visually inspected, or when the SDRi is below $0$, as this implies a decreasing performance over iterations, which is not desirable.}
    \label{fig:val_sdr}
\end{figure}

First, we study the impact of the step size on the performance of the proposed algorithm using the validation set. The mean SDRi on this subset is presented in Figure~\ref{fig:val_sdr} in the ``right" setting, but similar conclusions can be drawn in the ``left" setting. For $d=1$, we remark that the range of possible step sizes becomes more limited as $\beta$ decreases towards $0$ (which corresponds to the IS divergence). Conversely, when $d=2$, we observe that divergences corresponding to $\beta$ close to $1$ (i.e., the KL divergence) allow for more flexibility when it comes to choosing an appropriate step size.

\noindent For each setting, we pick the value of the step size that maximizes the SDR on this subset and use it in the following experiment.

\subsection{Comparison to other methods}

\begin{figure}[t]
    \centering
    \includegraphics[width=0.8\linewidth]{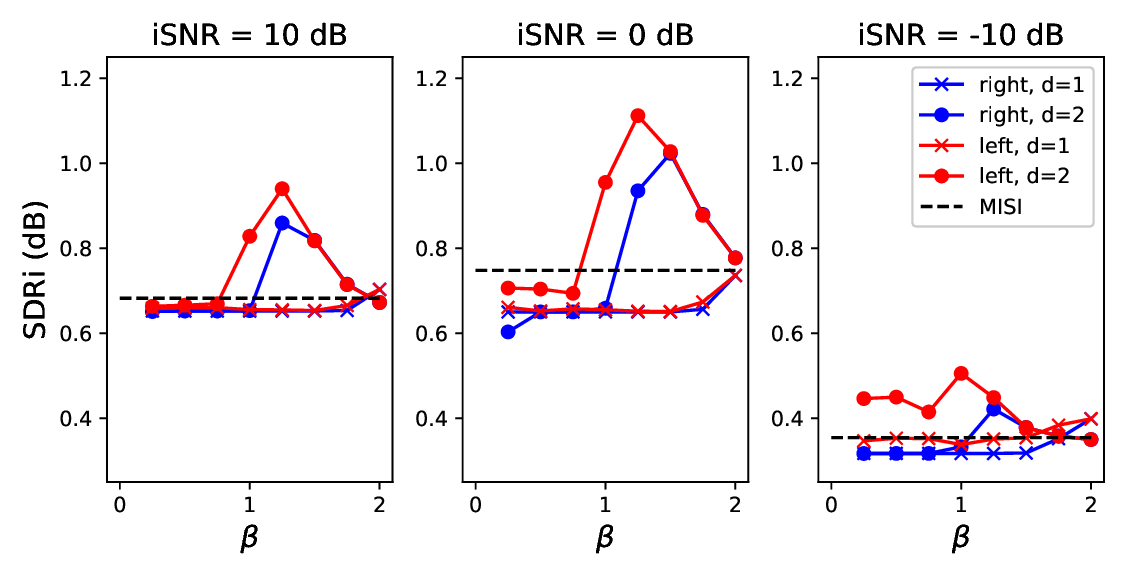}
    \caption{Average SDRi on the test set obtained with MISI and with the proposed algorithm (in different settings) at various iSNRs.}
    \label{fig:test_sdr}
\end{figure}

The separation results on the test set are presented in Figure~\ref{fig:test_sdr}. We observe that at high ($10$ dB) or moderate ($0$ dB) iSNRs, the proposed algorithm overall outperforms MISI when $d=2$ and for $\beta \geq 1$. We notably remark a performance peak at around $\beta = 1.25$ depending on the iSNR. This observation is consistent with the findings of~\cite{Vial2021}, where the gradient algorithm using the KL divergence (i.e., $\beta=1$) in a similar scenario ($d=2$ and ``left" formulation) exhibited good performance.

At low iSNR ($-10$ dB), the proposed method outperforms the MISI baseline when $d=2$ and for the ``left" problem formulation. This behavior is somewhat reminiscent of~\cite{Vial2021}: when the spectrograms are severly degraded (i.e., at low iSNR), the algorithm based on the quadratic loss (here, MISI) is outperformed by algorithms based on more suitable alternative losses.
Besides, it is also outperformed by a gradient algorithm based on the same quadratic loss when using a fine-tuned step size. This highlights the potential interest of phase recovery with Bregman divergences in such a scenario.

Finally, note that the performance of the proposed method strongly depends on the speaker and the kind of noise used in the experiments (noise-specific results can be obtained using the provided code, but we do not detail them here due to space constraints). For instance, for public square and bus noises, the proposed method consistently outperforms MISI at $10$ dB iSNR while both methods perform similarly at $-10$ dB iSNR. However, for living room noises, a different trend is observed: in particular, the improvement of the proposed algorithm over MISI becomes more significant at $-10$ dB iSNR.
As a result, further investigations are needed to identify the optimal $\beta$ for a given class of signals, which should reduce this sensitivity and improve the above results.

\section{Conclusion}
\label{sec:conc}

In this paper, we have addressed the problem of phase recovery with Bregman divergences for audio source separation. We derived a projected gradient algorithm for optimizing the resulting loss. We experimentally observed that when the spectrograms are highly degraded, some of these Bregman divergences induce better speech enhancement performance than the quadratic loss, upon which the widely-used MISI algorithm builds.

In future work, we will explore other optimization schemes for addressing this problem, such as majorization-minimization or ADMM, which has recently been introduced for tackling a similar problem using KL and IS divergences~\cite{Masuyama2020}. We will also leverage these algorithms in a deep unfolding paradigm for end-to-end and time-domain source separation.

\newpage
\bibliographystyle{IEEEbib}
\bibliography{references}

\begin{thebibliography}{10}

\bibitem{Comon2010}
P.~Comon and C.~Jutten,
\newblock {\em Handbook of blind source separation: independent component
  analysis and applications},
\newblock Academic press, 2010.

\bibitem{Barker2018}
J.~Barker, S.~Watanabe, E.~Vincent, and J.~Trmal,
\newblock ``The fifth '{CHiME}' speech separation and recognition challenge:
  Dataset, task and baselines,''
\newblock in {\em Proc. Interspeech}, Sept. 2018.

\bibitem{Cano2019}
E.~{Cano}, D.~{FitzGerald}, A.~{Liutkus}, M.~D. {Plumbley}, and F.~{Stöter},
\newblock ``Musical source separation: An introduction,''
\newblock {\em IEEE Signal Processing Magazine}, vol. 36, no. 1, pp. 31--40,
  Jan. 2019.

\bibitem{Wang2018c}
D.~{Wang} and J.~{Chen},
\newblock ``Supervised speech separation based on deep learning: An overview,''
\newblock {\em IEEE/ACM Transactions on Audio, Speech, and Language
  Processing}, vol. 26, no. 10, pp. 1702--1726, Oct. 2018.

\bibitem{Luo2019}
Y.~{Luo} and N.~{Mesgarani},
\newblock ``{Conv-TasNet}: Surpassing ideal time–frequency magnitude masking
  for speech separation,''
\newblock {\em IEEE/ACM Transactions on Audio, Speech, and Language
  Processing}, vol. 27, no. 8, pp. 1256--1266, Aug. 2019.

\bibitem{Luo2020}
Y.~{Luo}, Z.~Chen, and T.~Yoshioka,
\newblock ``Dual-path {RNN}: efficient long sequence modeling for time-domain
  single-channel speech separation,''
\newblock in {\em Proc. IEEE International Conference on Acoustics, Speech and
  Signal Processing (ICASSP)}, Apr. 2020.

\bibitem{Ditter2020}
D.~{Ditter} and T.~{Gerkmann},
\newblock ``A multi-phase gammatone filterbank for speech separation via
  {T}asnet,''
\newblock in {\em Proc. IEEE International Conference on Acoustics, Speech and
  Signal Processing (ICASSP)}, Apr. 2020.

\bibitem{Magron2018b}
P.~Magron, K.~Drossos, S.~I. Mimilakis, and T.~Virtanen,
\newblock ``{Reducing interference with phase recovery in DNN-based monaural
  singing voice separation},''
\newblock in {\em {Proc. Interspeech}}, Sept. 2018.

\bibitem{Gerkmann2015}
T.~Gerkmann, M.~Krawczyk-Becker, and J.~{Le Roux},
\newblock ``Phase processing for single-channel speech enhancement: History and
  recent advances,''
\newblock {\em IEEE Signal Processing Magazine}, vol. 32, no. 2, pp. 55--66,
  Mar. 2015.

\bibitem{Wang2018a}
Z.-Q. Wang, J.~{Le Roux}, D.~Wang, and J.~R. Hershey,
\newblock ``End-to-end speech separation with unfolded iterative phase
  reconstruction,''
\newblock in {\em {Proc. Interspeech}}, Sept. 2018.

\bibitem{Wichern2018}
G.~Wichern and J.~{Le Roux},
\newblock ``{Phase reconstruction with learned time-frequency representations
  for single-channel speech separation},''
\newblock in {\em {Proc. International Workshop on Acoustic Signal Enhancement
  (IWAENC)}}, Sept. 2018.

\bibitem{Wisdom2019}
S.~{Wisdom}, J.~R. {Hershey}, K.~{Wilson}, J.~{Thorpe}, M.~{Chinen},
  B.~{Patton}, and R.~A. {Saurous},
\newblock ``Differentiable consistency constraints for improved deep speech
  enhancement,''
\newblock in {\em Proc. IEEE International Conference on Acoustics, Speech and
  Signal Processing (ICASSP)}, May 2019.

\bibitem{Gunawan2010}
D.~Gunawan and D.~Sen,
\newblock ``Iterative phase estimation for the synthesis of separated sources
  from single-channel mixtures,''
\newblock {\em IEEE Signal Processing Letters}, vol. 17, no. 5, pp. 421--424,
  May 2010.

\bibitem{Gray1980}
R.~Gray, A.~Buzo, A.~Gray, and Y.~Matsuyama,
\newblock ``Distortion measures for speech processing,''
\newblock {\em IEEE Transactions on Acoustics, Speech, and Signal Processing},
  vol. 28, no. 4, pp. 367--376, Aug. 1980.

\bibitem{Hennequin2011}
R.~{Hennequin}, B.~{David}, and R.~{Badeau},
\newblock ``Beta-divergence as a subclass of {B}regman divergence,''
\newblock {\em IEEE Signal Processing Letters}, vol. 18, no. 2, pp. 83--86,
  Feb. 2011.

\bibitem{Smaragdis2014}
P.~Smaragdis, C.~F{\'e}votte, G.~J. Mysore, N.~Mohammadiha, and M.~Hoffman,
\newblock ``Static and dynamic source separation using nonnegative
  factorizations: A unified view,''
\newblock {\em IEEE Signal Processing Magazine}, vol. 31, no. 3, pp. 66--75,
  May 2014.

\bibitem{Vial2021}
P.-H. Vial, P.~Magron, T.~Oberlin, and C.~Févotte,
\newblock ``Phase retrieval with bregman divergences and application to audio
  signal recovery,''
\newblock {\em IEEE Journal of Selected Topics in Signal Processing}, vol. 15,
  no. 1, pp. 51--64, Jan. 2021.

\bibitem{Combettes2011}
P.~L. Combettes and J.-C. Pesquet,
\newblock ``Proximal splitting methods in signal processing,''
\newblock in {\em Fixed-point algorithms for inverse problems in science and
  engineering}, pp. 185--212. Springer, 2011.

\bibitem{Griffin1984}
D.~Griffin and J.~S. Lim,
\newblock ``{Signal estimation from modified short-time {F}ourier transform},''
\newblock {\em IEEE Transactions on Acoustics, Speech and Signal Processing},
  vol. 32, no. 2, pp. 236--243, April 1984.

\bibitem{Qiu2016}
T.~Qiu, P.~Babu, and D.~P. Palomar,
\newblock ``{PRIME}: Phase retrieval via majorization-minimization,''
\newblock {\em IEEE Transactions on Signal Processing}, vol. 64, no. 19, pp.
  5174--5186, Oct. 2016.

\bibitem{Perraudin2013}
N.~Perraudin, P.~Balazs, and P.~L. Sondergaard,
\newblock ``{A fast {Griffin-Lim} algorithm},''
\newblock in {\em {Proc. IEEE Workshop on Applications of Signal Processing to
  Audio and Acoustics (WASPAA)}}, Oct. 2013.

\bibitem{Zhu2007}
X.~Zhu, G.~T. Beauregard, and L.~L. Wyse,
\newblock ``Real-time signal estimation from modified short-time {Fourier}
  transform magnitude spectra,''
\newblock {\em IEEE Transactions on Audio, Speech, and Language Processing},
  vol. 15, no. 5, pp. 1645--1653, 2007.

\bibitem{Magron2020}
P.~Magron and T.~Virtanen,
\newblock ``Online spectrogram inversion for low-latency audio source
  separation,''
\newblock {\em IEEE Signal Processing Letters}, vol. 27, pp. 306--310, 2020.

\bibitem{Magnus1985}
J.~R. Magnus and H.~Neudecker,
\newblock ``Matrix differential calculus with applications to simple,
  {Hadamard}, and {K}ronecker products,''
\newblock {\em Journal of Mathematical Psychology}, vol. 29, pp. 474--492, Dec.
  1985.

\bibitem{Smith2011}
J.~O. Smith,
\newblock {\em Spectral audio signal processing},
\newblock W3K publishing, 2011.

\bibitem{Valentini2017}
{C. Valentini-Botinhao},
\newblock ``Noisy speech database for training speech enhancement algorithms
  and {TTS} models,'' \url{https://doi.org/10.7488/ds/2117}, 2017.

\bibitem{Thiemann2013}
J.~Thiemann, N.~Ito, and E.~Vincent,
\newblock ``{DEMAND: a collection of multi-channel recordings of acoustic noise
  in diverse environments},'' \url{https://doi.org/10.5281/zenodo.1227121},
  June 2013.

\bibitem{Stoter2019}
F.-R. St{\"o}ter, S.~Uhlich, A.~Liutkus, and Y.~Mitsufuji,
\newblock ``Open-{U}nmix - a reference implementation for music source
  separation,''
\newblock {\em Journal of Open Source Software}, 2019.

\bibitem{Valentini2016}
C.~Valentini-Botinhao, X.~Wang, S.~Takaki, and J.~Yamagishi,
\newblock ``Speech enhancement for a noise-robust text-to-speech synthesis
  system using deep recurrent neural networks,''
\newblock in {\em {Proc. Interspeech}}, Sept. 2016.

\bibitem{Uhlich2020}
S.~Uhlich and Y.~Mitsufuji,
\newblock ``Open-unmix for speech enhancement {(UMX SE)},''
  \url{https://doi.org/10.5281/zenodo.3786908}, May 2020.

\bibitem{Masuyama2020}
Y.~{Masuyama}, K.~{Yatabe}, K.~{Nagatomo}, and Y.~{Oikawa},
\newblock ``Joint amplitude and phase refinement for monaural source
  separation,''
\newblock {\em IEEE Signal Processing Letters}, vol. 27, pp. 1939 -- 1943,
  2020.

\end{thebibliography}

\end{document}